\documentclass[12pt,thmsa]{article}
\usepackage{sw20lart}

\input{tcilatex}
\begin{document}

\author{Coutinho, FAB$^{1}$; Burattini,MN$^{1}$; Lopez, LF$^{1}$ and Massad, E$%
^{1,2} $ \\
%EndAName
$^{1}$School of Medicine, \\
The University of S\~{a}o Paulo \\
and \\
LIM 01/HCFMUSP\\
Av. Dr. Arnaldo, 455, \\
S\~{a}o Paulo\\
CEP 01246-903, SP. \\
Brazil.\\
$^{2}$London School of Hygiene and\\
Tropical Medicine, London University,\\
U.K.\\
edmassad@usp.br}
\title{An aproximate threshold condition for non-autonomous system: an application
to a vector-borne infection}
\maketitle

\begin{abstract}
An non-autonomous system is proposed to model the seasonal pattern of dengue
fever.

We found that an approximate threshold condition for infection persistence
describes all possible behavior of the system.

As far as we know, the kind of analysis here proposed is entirely new. No
precise mathematical theorems are demonstrated but we give enough numerical
evidence to support the conclusions.

\textbf{keywords:} non-autonomous systems, stability analysis, thresholds,
dengue, epidemics.
\end{abstract}

\section{Introduction}

In a previous paper \cite{nóis1}, in which we tried to understand the
phenomenon of dengue overwintering, we discovered an interesting threshold
condition that allows the complete qualitative understanding of the behavior
of a non-autonomous system.

To motivate the reader we briefly describe the phenomenon we studied in \cite
{nóis1}.

In subtropical regions dengue fever, a mosquito transmitted disease, shows a
resurgent pattern with yearly epidemics, which starts typically in the
months characterized by heavy rains and heat, peaking some three or four
months after the beginning of the rainy season. In the dry months the number
of cases drop essentially to zero due to the virtual disappearance of the
vector. Since the infection reappears for some years in the same regions, it
is natural to ask how the virus survives the dry season.

In order to model those seasonal patterns of the disease, we proposed a
non-autonomous system, described below.

The model describes the dynamic of dengue in its three components of
transmission, namely, human hosts, mosquitoes and their eggs (the latter
includes the intermediate stages, like larvae and pupae). These populations,
in turn, are divided into susceptible humans, denoted $S_{H}$, infected
humans, $I_{H}$, recovered (and immune) humans, $R_{H}$, total humans, $%
N_{H}=S_{H}+I_{H}+R_{H}$, susceptible mosquitoes, $S_{M}$, infected and
latent mosquitoes, $L_{M}$, infected and infectious mosquitoes, $I_{M}$,
non-infected eggs, $S_{E}$, and infected eggs, $I_{E}$.

The model's dynamics is described by the set of equations.

\begin{equation}
\begin{array}{lll}
&  &  \\ 
\frac{dS_{H}}{dt} & = & -abI_{M}\frac{S_{H}}{N_{H}}-\mu
_{H}S_{H}+r_{H}N_{H}(1-\frac{N_{H}}{k_{H}}) \\ 
&  &  \\ 
\frac{dI_{H}}{dt} & = & abI_{M}\frac{S_{H}}{N_{H}}-(\mu _{H}+\alpha
_{H}+\gamma _{H})I_{H} \\ 
&  &  \\ 
\frac{dR_{H}}{dt} & = & \gamma _{H}I_{H}-\mu _{H}R_{H} \\ 
&  &  \\ 
\frac{dS_{M}}{dt} & = & p_{S}\left( c_{S}-d_{S}\sin \left( 2\pi ft+\pi
\right) \right) S_{E}\theta \left( c_{S}-d_{S}\sin \left( 2\pi ft+\pi
\right) \right) \\ 
&  & -\mu _{M}S_{M}-aS_{M}\frac{I_{H}}{N_{H}} \\ 
&  &  \\ 
\frac{dL_{M}}{dt} & = & aS_{M}\frac{I_{H}}{N_{H}}-e^{-\mu _{M}\tau
_{I}}aS_{M}(t-\tau _{I})\frac{I_{H}(t-\tau _{I})}{N_{H}(t-\tau _{I})}-\mu
_{M}L_{M} \\ 
&  &  \\ 
\frac{dI_{M}}{dt} & = & e^{-\mu _{M}\tau _{I}}aS_{M}(t-\tau _{I})\frac{%
I_{H}(t-\tau _{I})}{N_{H}(t-\tau _{I})}-\mu _{M}I_{M}+ \\ 
&  & p_{I}\left( c_{I}-d_{I}\sin \left( 2\pi ft+\pi \right) \right)
I_{E}\theta \left( c_{I}-d_{I}\sin \left( 2\pi ft+\pi \right) \right) \\ 
&  &  \\ 
\frac{dS_{E}}{dt} & = & \left[ r_{M}S_{M}+\left( 1-g\right)
r_{M}I_{M}\right] \left( 1-\frac{(S_{E}+I_{E})}{k_{E}}\right) - \\ 
&  & \mu _{E}S_{E}-p_{S}\left( c_{S}-d_{S}\sin \left( 2\pi ft+\pi \right)
\right) S_{E}\theta \left( c_{E}-d_{E}\sin \left( 2\pi ft+\pi \right) \right)
\\ 
&  &  \\ 
\frac{dI_{E}}{dt} & = & gr_{M}I_{M}\left( 1-\frac{(S_{E}+I_{E})}{k_{E}}%
\right) -\mu _{E}I_{E}- \\ 
&  & p_{I}\left( c_{I}-d_{I}\sin \left( 2\pi ft+\pi \right) \right)
I_{E}\theta \left( c_{I}-d_{I}\sin \left( 2\pi ft+\pi \right) \right)
\end{array}
\label{1}
\end{equation}

Let us briefly describe some features of the model.

We begin by describing the first three equations of the model.

Susceptible humans grow at the rate $r_{H}N_{H}(1-\frac{N_{H}}{k_{H}})-\mu
_{H}S_{H}$, where $r_{H}$ is the birth rate, $\mu _{H}$ is the natural
mortality and $k_{H}$ is the human carrying capacity. Note that we are
assuming that close to the carrying capacity the human population growth is
checked by a reduction in the birth rate. Alternatively the control of the
population could be done by assuming an increase in the mortality rate, but
the net result would be qualitatively the same. Those susceptible humans who
acquire the infection do so at the rate $abI_{M}\frac{S_{H}}{N_{H}}$, where $%
a$ is the average daily biting rates of mosquitoes and $b$ is the fraction
of infective bites inflicted by infectious mosquitoes $I_{M}$. Infected
humans, $I_{H}$ may either recover, with rate $\gamma $, or die from the
disease, with rate $(\mu _{H}+\alpha _{H})$. Recovered humans remain so for
the rest of their lives.

The fourth, fifth and sixth equations represent the susceptible, latent and
infectious mosquitoes sub-populations, respectively. Susceptible mosquitoes
varies in size with a time-dependent rate 
\[
p_{S}\left( c_{S}-d_{S}\sin \left( 2\pi ft+\pi \right) \right) S_{E}\theta
\left( c_{S}-d_{S}\sin \left( 2\pi ft+\pi \right) \right) -\mu _{M}S_{M}. 
\]
The term $\mu _{M}$ is the natural mortality rate of mosquitoes. The term $%
p_{S}S_{E}$ is the fraction of eggs present at time $t$, and which survived
the development through the intermediate stages (larvas and pupas). The
time-dependent rate $\left( c_{i}-d_{i}\sin \left( 2\pi ft+\pi \right)
\right) $ $\theta \left( c_{i}-d_{i}\sin \left( 2\pi ft+\pi \right) \right) $
$(i=S,I)$ simulates the seasonal variation in mosquitoes production from
eggs, assumed different for infected and susceptible eggs, for generality.
By varying $c_{i}$ and $d_{i\text{ }},(i=S,I),$ we can simulate the duration
and severity of the winters ($f$ $=1/365$ days$^{-1}$ and so it fixes one
cycle per year). The Heaviside $\theta $-function (a step function that is
equal to zero when the argument is less than zero and one when the argument
is greater or equal to zero) $\theta \left( c_{i}-d_{i}\sin \left( 2\pi
ft+\pi \right) \right) $ prevents the term

$\left( c_{i}-d_{i}\sin \left( 2\pi ft+\pi \right) \right) $ $\theta \left(
c_{i}-d_{i}\sin \left( 2\pi ft+\pi \right) \right) $ $(i=S,I)$

from becoming negative. If $c_{i}$ is smaller than $d_{i}$, then the winter
is long and severe. On the other hand, if $c_{i}$ is greater than $d_{i}$,
then the winter is short and mild. Susceptible mosquitoes who acquire the
infection do so at the rate $aS_{M}\frac{I_{H}}{N_{H}}$, where $a$ is the
average daily biting rates of mosquitoes, and became latent. A fraction of
the latent mosquitoes survives the extrinsic incubation period with
probability $e^{-\mu _{M}\tau _{I}}$ and become infectious. Therefore, the
rate of mosquitos becoming infectious per unit time is $e^{-\mu _{M}\tau
_{I}}aS_{M}(t-\tau _{I})\frac{I_{H}(t-\tau _{I})}{N_{H}}$. The term 
\[
p_{I}\left( c_{I}-d_{I}\sin \left( 2\pi ft+\pi \right) \right) I_{E}\theta
\left( c_{I}-d_{I}\sin \left( 2\pi ft+\pi \right) \right) 
\]
represent vertical transmission, that is, the rate by which infected eggs
become infectious adults. Infected mosquitoes die at the same rate $\mu _{M}$
as the susceptible ones.

The seventh and eighth equations represent the dynamics of susceptible and
infected eggs, respectively.

In the seventh equation, the term

\[
\left[ r_{M}S_{M}+\left( 1-g\right) r_{M}I_{M}\right] \left( 1-\frac{%
(S_{E}+I_{E})}{k_{E}}\right) 
\]

represent the birth rate of susceptible eggs born from susceptible
mosquitoes with rate 
\[
r_{M}S_{M}\left( 1-\frac{(S_{E}+I_{E})}{k_{E}}\right) 
\]
and from a fraction $\left( 1-g\right) $ of infected mosquitoes, with rate 
\[
\left( 1-g\right) r_{M}I_{M}\left( 1-\frac{(S_{E}+I_{E})}{k_{E}}\right) 
\]

The term $r_{M}\left( 1-\frac{(S_{E}+I_{E})}{k_{E}}\right) $ represents the
density-dependent rate of eggs birth rate. Once again we choose a density
dependence on birth rather than on death. Alternatively the control of the
population could be done by assuming an increase in the mortality rate $\mu
_{E}$, but the net result would be qualitatively the same. Finally, in the
last equation the term 
\[
gr_{M}I_{M}\left( 1-\frac{(S_{E}+I_{E})}{k_{E}}\right) -\mu _{E}I_{E} 
\]

represents the rate by which infected eggs grow and the term 
\[
p_{I}\left( c_{I}-d_{I}\sin \left( 2\pi ft+\pi \right) \right) I_{E}\theta
\left( c_{I}-d_{I}\sin \left( 2\pi ft+\pi \right) \right) , 
\]

as already mentioned, is the fraction of the hatched infected eggs which
evolves to infectious adults.

\section{An approximated threshold condition}

In the first part of this section we deduce a threshold condition for
epidemic. The intuition behind the procedures is discussed later on.

In order to deduce the threshold condition for epidemic we replace the
non-autonomous system (\ref{1}) by a autonomous one, by regarding the time
on the right side of the system (\ref{1}) as a parameter and then carry out
a local stability analysis. We linearize the second, the fifth, the sixth
and eighth equations of the autonomous system around a small amount of
disease $i_{H},$ $l_{M}$, $i_{M}$ and $i_{E}$: 
\begin{equation}
\begin{array}{lll}
\frac{di_{H}}{dt} & = & ab\frac{S_{H}}{N_{H}}i_{M}-(\mu _{H}+\alpha
_{H}+\gamma _{H})i_{H} \\ 
&  &  \\ 
\frac{dl_{M}}{dt} & = & a\frac{S_{M}}{N_{H}}i_{H}-\mu _{M}l_{M} \\ 
&  & -e^{-\mu _{M}\tau _{I}}a\frac{N_{M}(t-\tau _{I})}{N_{H}(t-\tau _{I})}%
i_{H}(t-\tau _{I}) \\ 
&  &  \\ 
\frac{di_{M}}{dt} & = & e^{-\mu _{M}\tau _{I}}a\frac{N_{M}(t-\tau _{I})}{%
N_{H}(t-\tau _{I})}i_{H}(t-\tau _{I})-\mu _{M}i_{M}+ \\ 
&  & p_{I}\left( c_{I}-d_{I}\sin \left( \Phi \right) \right) i_{E}\theta
\left( c_{I}-d_{I}\sin \left( \Phi \right) \right)  \\ 
&  &  \\ 
\frac{di_{E}}{dt} & = & gr_{M}\left( 1-\frac{(S_{E})}{k_{E}}\right)
i_{M}-\mu _{E}i_{E}- \\ 
&  & p_{I}\left( c_{I}-d_{I}\sin \left( \Phi \right) \right) i_{E}\theta
\left( c_{I}-d_{I}\sin \left( \Phi \right) \right) 
\end{array}
\label{44}
\end{equation}
where $\Phi =2\pi ft+\pi $.

We then examine the stability of the trivial solution of system (\ref{44}),
that is, $i_{E}=0$, $l_{M}=0$, $i_{H}=0$ and $i_{M}=0,$ as if the system
were autonomous\cite{els}. For this we assume the solutions: 
\begin{equation}
\begin{array}{lll}
i_{H} & = & c_{1}\exp \left( \lambda t\right) \\ 
&  &  \\ 
l_{M} & = & c_{2}\exp \left( \lambda t\right) \\ 
&  &  \\ 
i_{M} & = & c_{3}\exp \left( \lambda t\right) \\ 
&  &  \\ 
i_{E} & = & c_{4}\exp \left( \lambda t\right)
\end{array}
\label{444}
\end{equation}
drop the Heaviside $\theta -$functions by assuming $c_{I}\geq d_{I}$, and
replace (\ref{444}) into equation (\ref{44}). The characteristic equation
associated to system (\ref{44}) is then obtained:

\begin{equation}
\left| 
\begin{array}{llll}
-(\lambda +\gamma _{H}+\alpha _{H}+\mu _{H}) & 0 & ab\frac{S_{H}(t)}{N_{H}(t)%
} & 0 \\ 
&  &  &  \\ 
\begin{array}{l}
a\frac{S_{M}}{N_{H}}-ae^{\left( -\mu _{M}\tau \right) }\times  \\ 
\frac{N_{m}(t-\tau _{I})}{N_{H}(t-\tau _{I})}e^{-\lambda \tau }
\end{array}
& -(\lambda +\mu _{M}) & 0 & 0 \\ 
&  &  &  \\ 
ae^{\left( -\mu _{M}\tau \right) }\frac{N_{m}(t-\tau _{I})}{N_{H}(t-\tau
_{I})}e^{-\lambda \tau } & 0 & -(\lambda +\mu _{M}) & p_{I}\left(
c_{I}-d_{I}\sin \Phi \right)  \\ 
&  &  &  \\ 
0 & 0 & gr_{M}\left( 1-\frac{S_{E}}{k_{E}}\right)  & 
\begin{array}{l}
-\lambda -\mu _{E}- \\ 
p_{I}\left( c_{I}-d_{I}\sin \left( \Phi \right) \right) 
\end{array}
\end{array}
\right| =0  \label{44a}
\end{equation}

If all the roots of equation (\ref{44a}) have negative real parts, then the
equilibrium without disease is stable, that is, the origin is an atractor..
As shown in \cite{compte}, the first root that crosses the imaginary axis do
so through the real axis and this happens when

\begin{equation}
\begin{array}{lll}
R(t) & = & \frac{a}{(\gamma _{H}+\alpha _{H}+\mu _{H})}\frac{N_{m}(t-\tau
_{I})}{N_{H}(t-\tau _{I})}\frac{a\exp \left( -\mu _{M}\tau \right) bc}{\mu
_{M}}\frac{S_{H}(t)}{N_{H}(t)} \\ 
&  & + \\ 
&  & \frac{p_{I}\left( c_{I}-d_{I}\sin \Phi \right) gr_{M}\left( 1-\frac{%
S_{E}}{k_{E}}\right) }{\mu _{M}\left( \mu _{E}+p_{I}\left( c_{I}-d_{I}\sin
\Phi \right) \right) }>1
\end{array}
\label{4a}
\end{equation}

\bigskip

\smallskip Note that the first term in equation (\ref{4a}) is exactly the
expression proposed in \cite{mac} for the so-called 'basic reproduction
number'.

The intuition behind the above procedure is the following. System (\ref{1})
has 'no-mass', that is, it responds to perturbations instantaneously.
Therefore, we can find the time $t$ at which the stability of the trivial
solution of system (\ref{44}), that is, $i_{E}=0$, $i_{H}=0$ and $i_{M}=0$
becomes unstable. We have numerically checked that the time $t$ at which the
trivial solution (no-disease) of the autonomous system becomes unstable ($%
R>1 $) corresponds approximately to the moment at which the epidemic takes
off, that is, when the epidemic in system (\ref{1}) begins to increase as a
result of the introduction of a small amount of disease at time $t=0$.

\section{Qualitative analysis of the system's behaviors}

In this section we analyze qualitatively all the possible behaviors of the
system when a small amount of disease is introduced into a previously
uninfected population and when $R(t)$ is in its minimum value (winter time).
We do so by using $R(t)$ as in equation (\ref{4a}) and by numerically
simulating system (\ref{44}) with parameters values as in table 1.

\begin{center}
Table 1
\end{center}

We analyze two epidemiological scenarios, one in which $R(t)$, in the
absence of infection, is most of the time above one, and another in which $%
R(t)$ is most of the time below one. In the first case, if a small amount of
infection is introduced we observe a pattern shown in figure 1. In the
second case a small amount of infection introduced generates a pattern shown
in figure 2.

\begin{center}
Figure 1

Figure 2
\end{center}

In figure 1 the intensity of transmission is relatively low ($a=3.7$ days$%
^{-1}$) and we see a first peak followed by a succession of outbreaks
forming a damped oscillation pattern and the disease disappears. In other
words, after the first outbreak the infection transmission decreases to
levels inferior to that of the previous cycle. As the system oscillates and
the time interval during which $R(t)$ is above the threshold for
transmission is insufficient for keeping transmission, we have the pattern
observed.

In figure 2 the intensity of transmission is higher ($a=4.3$ days$^{-1}$)
than that shown in figure 1 and, consequently the time interval during which
the system is above the threshold is larger. In this case, the amplitude of
the consecutive outbreaks increases until the fraction of immune individuals
reaches a herd immunity threshold and the disease dies out in a damped
oscillation pattern.

We have numerically found that there is an increase in the amplitude of
consecutive outbreaks with $a=4.3$ days$^{-1},$ whenever the preceding
period of time $R(t)$ is above the threshold for transmission is greater
than a certain time interval (about 190 days). Indeed, with $a=3.7$ days$%
^{-1}$, as mentioned above, there is a first outbreak and the subsequent
peaks formed a damped oscilation and the time interval $R(t)$ is above the
threshold for transmission is greater than 190 days only for the first peak.

When we simulate the system with parameters that make $R(t)<1$ the disease
cannot invade the population and disappears exponentially. 

Those are the only three possible qualitative patterns generated by a small
amount of disease introduced into an entirely susceptible population when $%
R(t)$ is at its minimum value. 

Let us concentrate in the first two patterns, which can better be visualized
in figures 3 and 4, where the threshold parameter $R(t)$ is shown as
function of time for each of the above cases.

\begin{center}
Figure 3

Figure 4
\end{center}

Other interesting results are shown in figures 5 and 6, in which the time
oscillation of the `total amount of disease', $d(t)$, defined as 
\begin{equation}
d(t)=\sqrt{\left( I_{H}(t)\right) ^{2}+\left( L_{M}(t)\right) ^{2}+\left(
I_{M}(t)\right) ^{2}+\left( I_{E}(t)\right) ^{2}}  \label{5}
\end{equation}
is plotted together with $R(t)$, for both cases of low and high intensities
of transmission. It can be noted from the figure, that the points in which $%
R(t)$ crosses $1$ corresponds, approximately, to maximums and minimums of
the function $d(t)$. Note also that, in both cases the peaks and troughs of $%
d(t)$ occur slightly after $R(t)$ crosses $1$, decreasing and increasing,
respectively, as if the system has a small `inertia'.

\begin{center}
Figure 5

Figure 6
\end{center}

\section{Final comments}

This paper presents a novel, as far as we know, approach to analyze the
response of a non-autonomous system to a perturbation. This is quantified by
an approximate expression for the threshold condition that determines
whether the system will amplify or reduce a small of disease introduced at
time $t$. We should warn the reader that we have no mathematical proof of
the correctness of the threshold expression here deduced but the numerical
investigation presented above suggests that it is basically correct.

A possible application exemplified in this paper is the case of dengue
fever, in which a seasonal variation in the density of vector mosquitoes
determines the intensity of transmission. In another paper we used a similar
model to explain the question of overwintering, that is, how dengue fever
survives through the winter\'{}s dry and cold season.

Finally, we think that the analysis proposed in this paper could be applied
to other vector-borne infections and also to some directly transmitted
diseases that show seasonality in the intensity of transmission.

\section{Acknowledgments}

This work was supported by grants LIM01/HCFMUSP, CNPq, FAPESP and PRONEX.

\newpage

\begin{center}
\textbf{Captions for the Figures}
\end{center}

\textbf{Figure 1}. Number of infected humans for the case of low intensity
of transmission ($a=3.7$ days$^{-1}$). There is a first outbreak resulting
from the introduction of a small amount of infection in a previously
uninfected population followed by a pattern of damped oscillation until the
disease disappear. Simulation begins at the peak of the winter.

\textbf{Figure 2}. Number of infected humans for the case of high intensity
of transmission ($a=4.3$ days$^{-1}$). After the first outbreak resulting
from the introduction of a small amount of infection in a previously
uninfected population there are subsequent outbreaks with larger amplitudes
until herd immunity is achieved and the disease gradually disapears.
Simulation begins also at the peak of the winter.

\textbf{Figure 3}. The threshold $R(t)$ in the case of low transmission ($%
a=3.7$ days$^{-1}$).

\textbf{Figure 4}. The threshold $R(t)$ in the case of high transmission ($%
a=4.3$ days$^{-1}$).

\textbf{Figure 5}. The threshold $R(t)$ and `total amount of disease' $d(t)$
in the case of low transmission ($a=3.7$ days$^{-1}$). The peaks and troughs
of $d(t)$ occur slightly after $R(t)$ crosses $1.$

\textbf{Figure 6}. The threshold $R(t)$ and `total amount of disease' $d(t)$
in the case of high transmission ($a=4.3$ days$^{-1}$). Again, the peaks and
troughs of $d(t)$ occur slightly after $R(t)$ crosses $1.$

\newpage 

\begin{tabular}{lll}
Table 1 &  &  \\ 
\textbf{Parameter} & \textbf{Meaning} & \textbf{Value} \\ 
$a$ & Average Daily biting rate & see text \\ 
$b$ & Susceptibility to Infection & 0.1 \\ 
$\mu _{H}$ & Humans Natural Mortality Rate & 4 x 10$^{-5}$days$^{-1}$ \\ 
$r_{H}$ & Humans Malthusian Parameter & 1 days$^{-1}$ \\ 
$k_{H}$ & Humans Carrying Capacity & 10$^{6}$ \\ 
$\alpha _{H}$ & Dengue Induced Mortality in Humans & 10$^{-3}$days$^{-1}$ \\ 
$\gamma _{H}$ & Humans Recovery Rate & 0.143 days$^{-1}$ \\ 
$p_{S}$ & Proportion of non-infected eggs that reach adult phase & 0.15 \\ 
$c_{S}$ & Climatic factor modulating winters and summers & 0.08 \\ 
$d_{S}$ & Climatic factor modulating winters and summers & 0.06 \\ 
$f$ & Frequency of the seasonal cycles & 2.8 x 10$^{-3}$days$^{-1}$ \\ 
$\mu _{M}$ & Natural mortality rate of mosquitoes & 0.263 days$^{-1}$ \\ 
$\tau $ & Extrinsic incubation period of dengue & 7 days \\ 
$\alpha _{M}$ & Dengue induced mortality in mosquitoes & negligible \\ 
$r_{M}$ & Eggs Malthusian parameter & 50 days$^{-1}$ \\ 
$p_{I}$ & Proportion of infected eggs that reach adult phase & 0.15 \\ 
$c_{I}$ & Climatic factor modulating winters and summers & 0.06 \\ 
$d_{I}$ & Climatic factor modulating winters and summers & 0.06 \\ 
$g$ & Proportion of infected eggs laid by infected females & see text \\ 
$k_{E}$ & Eggs Carrying Capacity & 10$^{6}$ \\ 
$\mu _{E}$ & Natural mortality rate of eggs & 0.1 days$^{-1}$%
\end{tabular}

\end{document}